\documentclass[referee]{raa}           
\usepackage{graphicx,times}
\usepackage{natbib}
\usepackage{amssymb,amsmath}
\bibpunct{(}{)}{;}{a}{}{,}

\usepackage[a4paper=true,dvipdfm=true,pagebackref=true]{hyperref}
\hypersetup{pdftitle = The title of my PDF, pdfauthor = My name, pdfsubject= The subject, pdfkeywords = keyword1 keyword2 keyword3} 
\hypersetup{colorlinks = true, linkcolor = green, anchorcolor = red, citecolor = blue, filecolor = red, pagecolor = red, urlcolor = red}

\begin{document}


   \title{Fabry-P\'{e}rot based Narrow Band Imager for Solar Filament Observations}

 \volnopage{ {\bf 2015} Vol.\ {\bf X} No. {\bf XX}, 000--000}
   \setcounter{page}{1}

   \author{Sajal Kumar Dhara\inst{1}, Belur~Ravindra\inst{1}, Ravinder~Kumar~Banyal\inst{1}}
   

   \institute{ Indian Institute of Astrophysics, Bangalore-560034, India; {\it sajal@iiap.res.in}}

\abstract{
We have recently  developed a narrow band imager (NBI) using an air gap based Fabry-P\'{e}rot (FP) interferometer at the
Indian Institute of Astrophysics, Bangalore. 
Narrow band imaging is achieved by using an FP interferometer working in combination with an order sorting pre-filter. 
The NBI can be tuned to a different wavelength position on the line profile by changing the plate separation of the FP.
The interferometer has a 50~mm clear aperture with a bandpass of $\sim$247.8~m\AA~ and a free spectral range of 
$\sim$5.3~\AA~at $\lambda$~ = 656.3~nm.
The developed NBI is used to observe the solar filament in the H$\alpha$ wavelength.
The instrument is being used to image the Sun at chromospheric height and it is also able to scan the 
H$\alpha$ spectral line profile at different wavelength positions. 
We have also made Doppler velocity maps at chromospheric height by taking the 
blue and red wing images at $\pm$176 m\AA~ wavelength positions separately away from the line center of the
spectral line. In this paper, we present a description of 
the NBI including lab test results of individual components and  
some initial observations carried out with this instrument.
}

   \authorrunning{Dhara, Ravindra \& Banyal}            
   \titlerunning{Narrow Band Imager}  
   \maketitle


\section{Introduction}      
\label{s:intro}

It is well known that features in the solar atmosphere change very rapidly. During active 
phenomena \textit{e.g.} sunspots, filaments/prominences, flares etc., there is a rapid change
in the magnetic field and velocity field at various heights in the solar atmosphere. 
To understand solar activity on the Sun, it is important to 
measure the magnetic and velocity fields of the solar atmosphere at different heights. 
To measure the solar magnetic and velocity fields, a high resolution spectrograph or tunable narrow band imager (NBI)
is needed \citep{Zirin95}. Spectrograph based instruments can have a very high spectral 
resolution but to obtain a two-dimensional (2-D) image of the Sun,  spatial scanning is required 
across the desired field-of-view (FOV) \citep{Judge10}. Moreover the
FOV is also limited by the length and scan range of the spectrograph slit.
Using imaging spectroscopy one can obtain 2-D images of the Sun, but it requires scanning across multiple
wavelength positions to build up the spectra. It also suffers from low spectral resolution.
The rasterization of the spectrograph to make the 2-D image can be avoided in a narrow band imaging system.
In both cases, the ground based observations are affected by atmospheric turbulence.
By using high-order adaptive optics and/or post-facto techniques, one can minimize the atmospheric 
seeing effects to achieve high spatial resolution imaging over a limited FOV.

Narrow-band imaging can be carried out by a Fabry-P\'{e}rot (FP) interferometer. Using a single FP interferometer,
a narrow band imaging system can be 
developed. The spectral resolution can be further improved using two FP interferometers in tandem \citep{Cavallini06,Ravindra11}.  
Since the light throughput 
is very high for an FP based imaging system, one can obtain 2-D images by scanning across the 
spectral lines in shorter time periods depending upon the number of wavelength positions to be scanned along the line profile. 
One of the major advantages of this system is that it can be switched between wavelengths very fast. Hence one can 
obtain images in different wavelengths by combining the FP with a different set of pre-filters.
There are two types of FP interferometers; a) solid etalons and b) an air gap FP. In case of air gap FP, tuning is achieved by 
varying the separation between the two plates of the FP. However, for solid etalons 
the tuning is done  by varying the refractive index 
of the electro-optic material. To obtain 2-D images with a fast cadence both,
air gap FPs (\textit{e.g.,} \cite{Kentischer98,Cavallini06,Bello08}) as well as 
solid etalons (\textit{e.g.,} \cite{Rust96,Mathew98,Kleint11,Martinez11,Raja14}) have been used in imaging systems.
Air gap based FP requires a control system that maintains the parallelism
between the two highly polished plates to an accuracy of better than $\lambda$/100.
One disadvantage of this method is that it introduces a larger 
wavelength shift than electro-optically tunable solid-state etalons for the same FOV because of their lower refractive index.
It can also be operated at a faster rate as fast as LiNbO$_{3}$ etalons while tuning \citep{Martinez11}.

At the Indian Institute of Astrophysics (IIA), a narrow-band imaging system (imager) has
been developed using an air gap based FP interferometer in combination with an order sorting H$\alpha$ filter.
The FP has a 50~mm clear aperture with a 400 nm to 700 nm transmission wavelength range. 
The effective full width at half maximum (FWHM) and free spectral range of the FP are 238~m\AA~ and  5~\AA~at $\lambda$~ = 633~nm, respectively.
We have used an H$\alpha$ interference filter (central wavelength at 6562.8~\AA~ with FWHM 2.34~\AA~ at 30$^0$C) as an order sorting
pre-filter. In this paper, we briefly describe the characterization of the FP and order-sorting pre-filter and evaluate their
suitability for solar observations at the spectral line (H$\alpha$-6563~\AA). 

In Section ~\ref{s:Imager} we discuss
the schematic layout of the NBI, various tuning options and the observations made with this instrument.
In Section ~\ref{s:observation} we show some of the images taken with this instrument.
Discussions are provided in Section~\ref{s:discussion}.

\section{A brief description of the Instrument}\label{s:Imager}

\begin{figure}
\centerline{\includegraphics[width=0.8\textwidth,clip=]{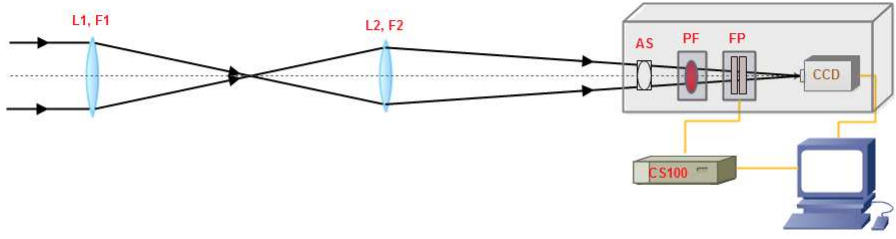}}
\caption{A schematic diagram representing the optical and control layout of the NBI. The abbreviations mean the following:
L1--15~cm objective lens with focal length F1=225 cm, L2--10 cm re-imaging lens with focal length F2=100 cm,
AS -- circular aperture stop, PF--pre-filter inside temperature 
controlled oven, FP--Fabry-P\'{e}rot interferometer and CCD--camera.\label{fig:1}}
\end{figure}

Fig.~\ref{fig:1} shows the schematic layout of the instrument. Light from the coelostat is imaged by the 
achromatic doublet lens (L1) with a 15 cm aperture at its focal point (F1). We have used a 10 cm aperture during observations
to reduce the intensity. The focal length of the objective lens
is 225 cm which makes an $F_{22.5}$ beam. The image size of the full Sun at the focal plane is 21 mm. Then it 
is re-imaged by another achromatic doublet lens (L2) with diameter and 
focal length of 10 cm and 100 cm, respectively. The lens L2 is placed at a distance of 130 cm away from the focal point
of the objective lens. Hence the final image is formed 433.3 cm away from lens L2. This produces an
$F_{75}$ beam at the imaging plane on the CCD. The image size of the Sun at this imaging plane is 70 mm.
A circular aperture stop (AS) is placed in front of the pre-filter (PF) to cut down the amount of scattered 
stray light inside the instrument box.
The main component of the NBI is the FP interferometer, which is 
mounted behind the pre-filter in the optical path. The details about FP and PF are given in section~\ref{s:fabry-perot} and 
section~\ref{s:prefilter}, respectively.

\subsection{Fabry-P\'{e}rot Interferometer}\label{s:fabry-perot}

An FP interferometer is a device which consists of two highly polished reflecting parallel glass plates. 
In combination with the order sorting filter, it selectively transmits a particular wavelength of 
light corresponding to the resonance of the etalon cavity. If the separation between the two plates of FP 
is \emph{d}, refractive index of the media between the plates $\mu$, and
wavelength of incident light  $\lambda$ with incident angle $\theta$ and order
\emph{m}, the interference intensity attains its maximum value when the following relationship is satisfied \citep{Born89}.

\begin{equation} \label{equ:1}
2\mu d cos{\theta} =m{\lambda}
\end{equation}

The intensity distribution of the transmitted beam is given by Airy's formula 

\begin{equation} \label{equ:2}
I = \frac{{I}_{max}}{1+F\sin^2(\frac{\delta}{2})}
\end{equation}

Where, $I_{max}=A^2T^2/(1$-$R)^2$ is the maximum value of the intensity and $F=4R/(1-R)^2$ is the finesse due to reflectivity
of the FP plates. T and R are the transmitivity and reflectivity of the FP respectively. A is the amplitude of the incident light. 
$\delta$ is the phase difference between the successive transmitted rays which is given by,

\begin{equation} \label{equ:3}
{\delta} = \frac{2\pi}{\lambda}(2\mu d cos{\theta})
\end{equation}

The phase difference can be varied by changing the refractive index of the material
between the plates or by changing the separation between the plates of the FP or by changing the
angle of incidence for the rays. The spectral line profile can be scanned by choosing one of the methods to vary the phase. 
The transmitted beam takes the form of periodic fringes with a maximum intensity of $I_{max}$ and the minimum intensity is
$I_{min}=A^2T^2/(1$+$R)^2$. 

The two key characteristics of FP are: (i) the finesse and
(ii) the free spectral range (FSR). The finesse is related to the sharpness of the interference fringes. 
It is the ratio between the distance of two neighboring peaks and their halfwidth. A measured finesse 
has a number of contributing factors: the mirror reflectivity, the mirror surface 
quality,  plate parallelism, etc. The FSR is the distance between two successive maxima
and is given by

\begin{equation} \label{equ:4}
FSR = \frac{\lambda}{m}=\frac{\lambda^2}{2\mu d cos{\theta}}
\end{equation}
Hence, the FSR is inversely proportional to the spacing between the plates.

We have an air gap based FP for the NBI system. This FP was procured from \textit{Queensgate Instruments} (now called 
\textit{IC Optical Systems Ltd}). 
The servo-stabilized FP system is comprised of an
ET-Series II etalon and the CS100 control unit, which stabilizes the etalon
spacing and parallelism. The CS100 is a three-channel controller, which uses
capacitance micrometers and PZT actuators, incorporated into the etalon, to
monitor and correct errors in mirror parallelism and spacing. Two
channels control the parallelism and the third channel maintains spacing up to nanometer accuracy by
referencing the cavity length-sensing capacitance micrometer to a
fixed reference capacitor. This is a closed-loop system. Hence, 
non-linearity, hysteresis in the PZT drive, drifts in mirror parallelism and spacing are completely eliminated.
The CS100 controls the etalon spacing and parallelism to better
than 0.01$\%$ of an FSR.
The CS100 can be operated manually from front panel controls, or under computer control using
either the IEEE - 488, RS232C or analogue interfaces. We have developed a GUI
in LabVIEW using the IEEE - 488 as well as the analog interface using PCIe - DAQ
to control the CS100 controller via PC. The separation between the plates is
varied in discrete steps from 0 to 4095 with the IEEE - 488 interface to the
CS100 controller via a PC. In addition to this, a two-pin socket is provided on the CS100 rear panel
to enable analog control of the etalon spacing. A $\pm$10V differential input can be
given using DAQ to produce $\pm$1000 nm of plate movement for the cavity scanning. 

This FP has a clear aperture of 50 mm and high average reflectivity ($\sim$95\%) for the wavelength range from 400 nm to
700 nm. The calculated finesse due to a reflectivity of 95\% is $\sim$61 and FSR is $\sim$ 5.0~\AA~ at $\lambda$= 633 nm.
The specifications of the FP system as provided by the vendor are listed in Table~\ref{Table:1}.
The performance characteristic of the FP needs to be checked against the specifications given by the vendor 
before being used in the NBI system. In the next subsection we briefly describe the methods used to evaluate the
performance characteristics of the FP and obtained results.

\begin{table}[h!]

\caption{Specification of ET-50 Fabry-P\'{e}rot system as provided by the vendor.} 
\centering

\begin{tabular}{|c|c|}

\hline

Parameter              & Value \\
\hline

\hline
Fabry-P\'{e}rot Model      & ET-50-FS(SN 1006)\\
\hline
Clear Aperture         & 50mm \\
\hline
Nominal Cavity Spacing & 400$\mu$m\\
\hline
Cavity Tuning Range    & $\pm$2.5$\mu$m\\
\hline
Reflectivity           & $\sim$95\% at $\lambda$=633nm\\
\hline
Transmission wavelength range & 400nm --700nm \\
\hline
Response Time when used with CS100 Controller & 0.2ms -- 2.0ms\\

\hline 
\end{tabular}
\label{Table:1}
\end{table}

\subsection{Characterization of FP}

\subsubsection{Experimental Setup}

Before using the procured FP interferometer in the NBI system, we evaluated its specification 
in laboratory experiments. 
We have characterized the FP using a collimated configuration. 
An He-Ne laser with a central wavelength of 633~nm was used as a source to perform the experiment. 
The beam was expanded and collimated using a spatial filter assembly and a lens system before passing through the FP. 
A 2k $\times$ 2k CCD, procured from \textit{Princeton Instruments}, was used to record the images. 
The CCD camera and spacing of FP were controlled by computer.

\subsubsection{Bandpass, free spectral range and finesse }

The laser line width is much narrower than the bandpass of the FP and was thus used as an 
impulse input to measure the FWHM of the transmission function of the FP. 
The transmitted images of the laser beam 
through the FP 
were obtained as the cavity spacing was varied from one maximum to another.

\begin{figure}    
  \centerline{\hspace*{0.015\textwidth}
               \includegraphics[width=0.45\textwidth,clip=]{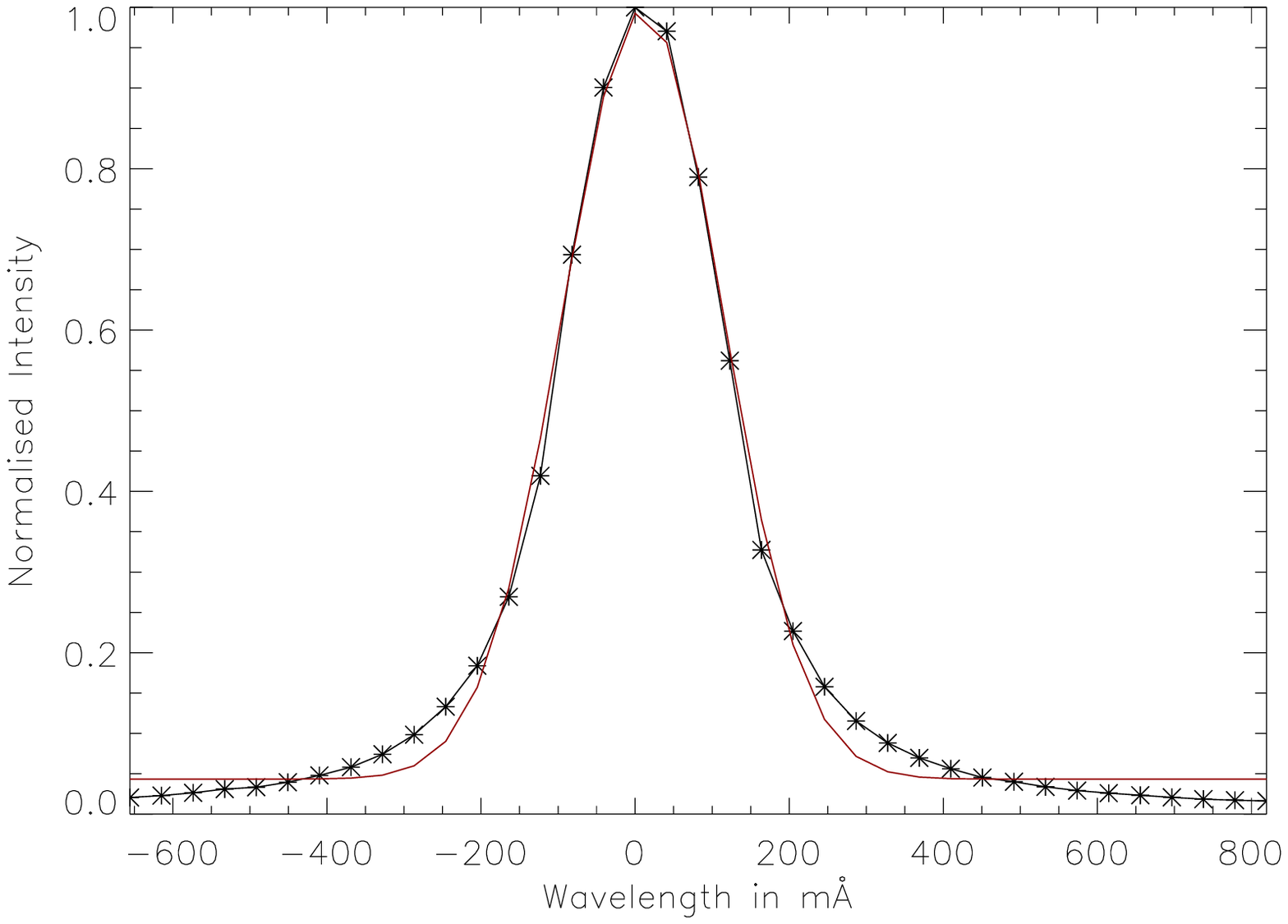}
               \hspace*{-0.02\textwidth}
               \includegraphics[width=0.45\textwidth,clip=]{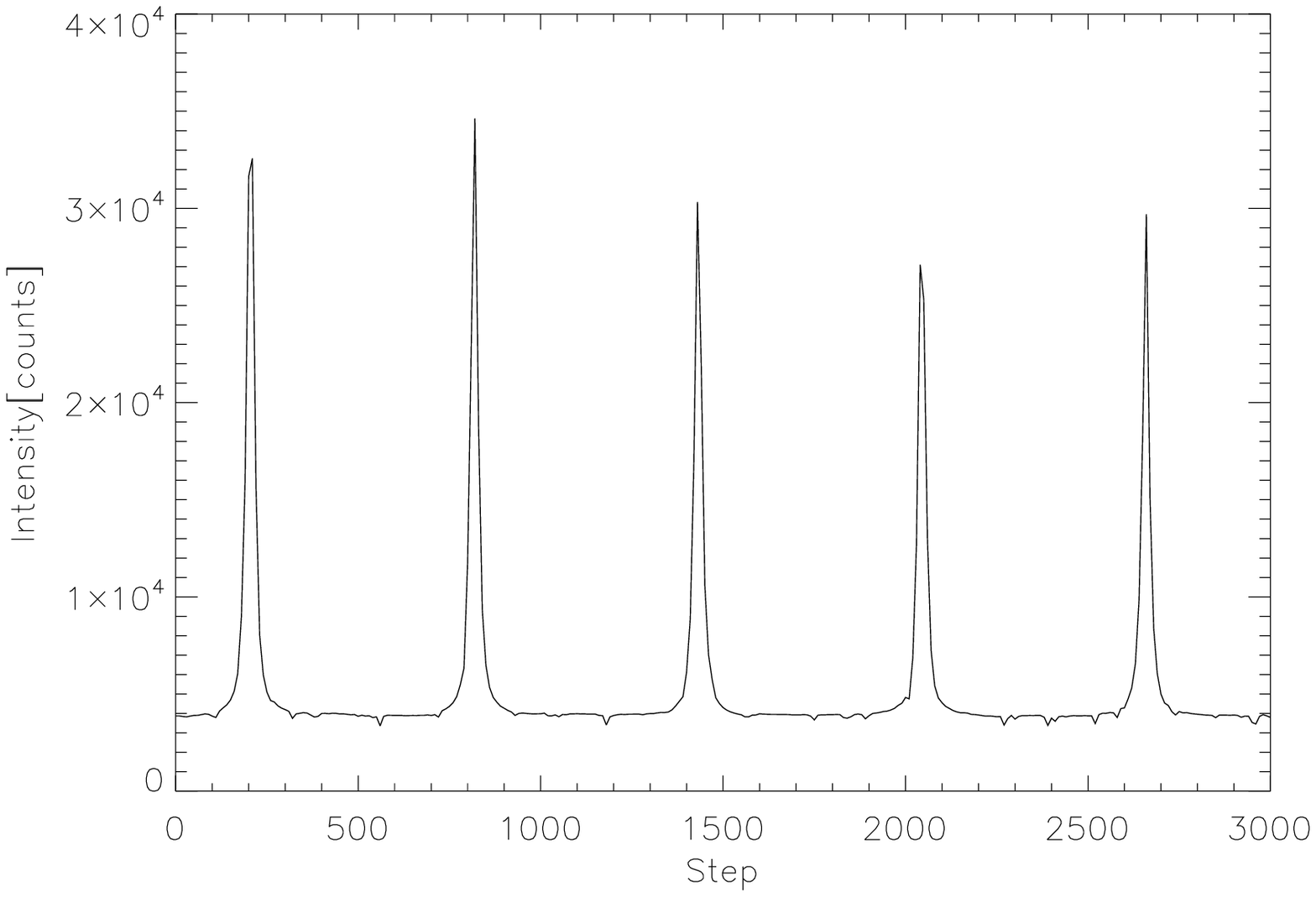}
              }
    
      \vspace{0.05\textwidth}    
\caption{Left: The average intensity profile of one transmission peak.
The red line shows a Gaussian fitting to the measured intensity profile.
Right: The average intensity profile of several transmission
peaks. The wavelength regime is given in discrete steps (0-3000) of
the CS100 controller.}

   \label{fig:fwhm_fsr}
   \end{figure}

An intensity profile, which is the average intensity in the central part of each of the 
transmitted images (100$\times$100 pixels$^2$) 
from the image data cube, is shown in Fig.~\ref{fig:fwhm_fsr} (left). 
The effective FWHM of the bandpass is obtained by fitting a Gaussian
profile of the measured intensity values. The effective FWHM is found  to be
$\sim$23.77$\pm$0.46 pm. 
If the scan range is extended further, more than one intensity peak can be obtained. Here, we carried out a scan from 0
to 4095 steps, which included five transmission peaks. 
Fig.~\ref{fig:fwhm_fsr} (right) shows the transmission peaks to illustrate the measurement of FSR of the FP. The
measured FSR is $\sim$499.93 pm. 

The average effective finesse is found to be 21.03. 
It is the most important part for
the performance evaluation of the FP interferometer. It is dependent on the finesse 
due to reflectivity ($F_{R}$), finesse due to plate defects ($F_{S}$, $F_{G}$ and $F_{P}$) and finesse due to 
 illumination of the plates ($F_{I}$). In detail, the finesse due to plate defects is dependent on overall flatness and 
roughness of the FP plates and the departure of the two plates from parallelism \citep{Gary03,Gullixson98}.

\begin{table}[h!]
\centering
\caption{Comparison between vendor supplied values and experimentally measured values.}
\begin{tabular}{|c|c|c|}
 \hline
 
Parameters & Vendor's specifications & Experimentally measured  \\
           &                        & values\\
\hline

Reflectivity                  &$\pm$95\%                              &$\pm$86\% \\
FSR                           &$\sim$5\AA($\lambda$=656.3 nm)         &$\sim$5\AA($\lambda$=633 nm) \\
FWHM                          &$\sim$81.7m\AA($\lambda$=656.3 nm)     &$\sim$237.7m\AA($\lambda$=633 nm) \\
Finesse due to Reflectivity   &$\sim$61.24                            &$\sim$21.38\\
Effective Finesse             &$\sim$61.73                            &$\sim$21.03\\
Surface Quality               & $\lambda$/50 - $\lambda$/200 at $\lambda$= 633 nm &  Flatness: $\sim$ $\lambda$/35, 
                                                                       Roughness:  $\sim$ $\lambda$/74 \\
\hline 
\end{tabular}
\label{Table:ch4_2}
\end{table}
A comparison between the vendor's specification and experimentally
measured values is given in Table~\ref{Table:ch4_2}.
We can conclude that the calculated effective finesse does not agree with the measured finesse. The following possible
reason might be helpful for interpreting the differences between calculated and measured values. 
a) The measured FWHM for this FP may have been broadened by the finite bandwidth of the laser beam,
resulting in a decrease of the measured effective finesse. 
b) If the expanded source is not perfectly collimated to feed the FP, it will result in lowering the finesse.
But before feeding the collimated beam to the FP, we used a shearing interferometer \citep{Darlin95,Zhao95}
and made sure that the beam was collimated.
We achieved the proper collimation using this method.
Hence, we can easily rule out this reason.
c) From the measured finesse, the estimated reflectivity of the FP plates is
about 86\%, which is not consistent with the vendor supplied values. 
Since our FP was purchased several years  ago, it is very likely that exposure to
humidity and dust may have caused the gradual degradation in reflectivity over time.

\subsubsection{FP Channel Spectra}
\label{FP_Channel_Spectra}
It can be noted that a collimated set-up is preferred for characterizing 
the FP. The reason is that the collimated set-up is theoretically insensitive
to any additional changes in optical path length caused due to insertion and removal of FPs
depending on whether the etalons are being characterized individually or in tandem. 
For obtaining the channel spectra of the FP, a Littrow spectrograph setup was used in 
conjunction with a telescope with a 15 cm clear aperture objective lens as the 
light feed. On the CCD, the dispersion per
pixel was 20.9 m\AA~ at 6563~\AA. 
We used 2k $\times$ 2k CCD camera, procured from \textit{Apogee Imaging Systems} to record the spectra. 

\begin{figure}
\begin{center}
\centerline{\hspace*{-0.1\textwidth}
\includegraphics[width=0.9\textwidth,clip=]{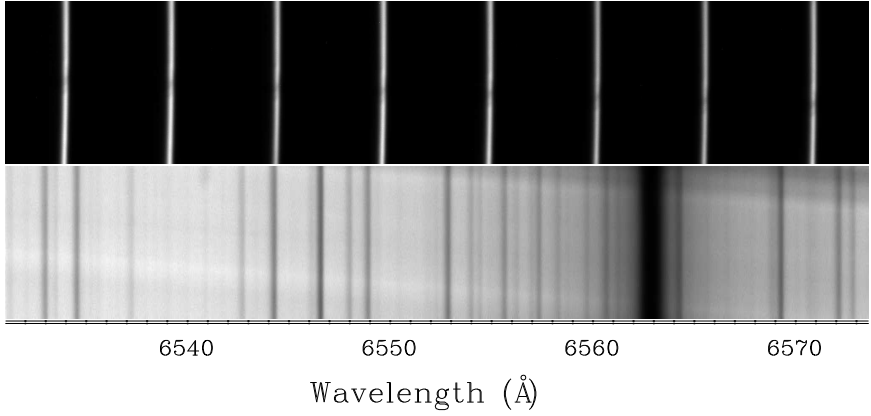}
           }
  \vspace{-0.1\textwidth}    
               
\end{center}
\caption{Top: Channel spectrum of FP. The intensity is expressed on a logarithmic scale.
Bottom: Observed solar spectrum.}
\label{fig:fp_channel_spectra}
\end{figure}

The FP was positioned into the collimated white light beam from the Sun.
Fig.~\ref{fig:fp_channel_spectra}~(top-panel) shows the channel spectrum obtained in this experiment. 
The bottom panel shows the solar spectrum which includes the 6563~\AA~ spectral line. 
The recorded channel spectrum was analyzed to obtain the FSR and the
FWHM for the FP. The measured FSR and FWHM from the transmission profile are
5.3~\AA~ and 320.2$\pm$3.2~m\AA~ respectively.
The FSR nearly matches the values obtained by the experiment done using a laser.
However, the FWHM measured with the channel spectrum is larger than the previously measured value. 
The theoretically calculated resolution of the spectrograph is 91~m\AA~ at 6563~\AA. 
The FWHM of the spectrograph $(\delta\lambda_{s})$ is estimated as follows: 

\begin{equation}
  \label{equ:6}
(\delta\lambda_{r})^{2} = (\delta\lambda_{s})^{2} + (\delta\lambda_{c})^{2}
\end{equation}

Where $(\delta\lambda_{r})$ is the FWHM of the recorded 6563~\AA~ spectral line profile and $(\delta\lambda_{c})$ is the 
obtained FWHM from the \textit{BASS2000} atlas. The FWHM of the spectrograph is estimated to be about 182.6~ m\AA.
This higher value could be due to the larger slit width (80~$\mu$m) and also due to the small F-ratio of the spectrograph
($F_{10}$), which covers
only $\frac{3}{4}$th of grating there by reducing the spectral resolution.


\subsection{Pre-filter}\label{s:prefilter}

 \begin{figure}    
        \centerline{\hspace*{-0.015\textwidth}
               \includegraphics[width=0.5\textwidth,clip=]{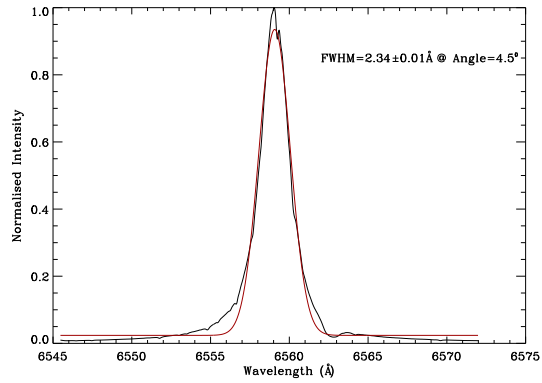}
              }
       \vspace{0.05\textwidth}    
     
\caption{Transmission profile of the filter at angle position of 4.5$^0$. 
The red line shows a Gaussian fitting to the measured intensity profile. 
}

   \label{fig:filter}
   \end{figure}
   

We use an H$\alpha$ filter centered at a wavelength of 6562.8~\AA~ as an order sorting filter placed in front of the FP.
It is an interference filter procured from \textit{Andover Corporation}. The central wavelength of the
filter gets shifted towards lower wavelength with an increase in the angle of incidence. Equation~\ref{equ:lambda}~ implies that the
central wavelength shift depends upon the angle of incidence and effective refractive index of the filter. 
The effective refractive index of the filter's dielectric material is 2.05. The specified temperature for the 
filter is 23$^0$C. 

\begin{equation}\label{equ:lambda}
{\lambda}_{\theta} = {\lambda}_{0} \sqrt{1 -\frac{\sin^2\theta}{n^2}}
\end{equation}

Where $\lambda_{\theta}$ and $\lambda_{0}$ are central wavelengths at an angle $\theta$ and zero, respectively, and \textit{n} is the
effective refractive index of the dielectric material of the filter. 

We developed a temperature stabilized box which can keep the filter within 0.1$^0$C of the set temperature in collaboration
with Udaipur Solar Observatory. The box temperature can be set between 30 and 50$^0$C with an accuracy of 0.1 $^{\circ}$C.
The stabilization circuit uses a temperature sensor IC (DS620) from \textit{Dallas Semiconductor}, which is a band-gap 
based temperature sensor. The temperature is read out using a computer parallel port and the correction to the signal 
is generated through Pulse Width Modulation and fed to the heater (see \cite{Gupta08} for details about the temperature controller).
  
We have characterized the filter using the spectrograph, as discussed in Section~\ref{FP_Channel_Spectra}. 
We kept the filter in the collimated beam ahead of the spectrograph slit.
The normalized filter transmission profile 
obtained at an angle position of 4.5$^0$ is shown  
in Fig.~\ref{fig:filter}. The tilt is to avoid the H$\alpha$ line entering the transmission profile of the 
filter. The tilting of the filter can increase the FWHM of the filter profile. However, the increase is only significant 
above 10$^{\circ}$ and below which it is well below 0.2~\AA, which is tolerable. 
The red line in the plot represents the Gaussian fit to the measured profile. The FWHM of the filter 
is found to be 2.34$\pm$0.01~\AA.

\subsection{Fabry-P\'{e}rot interferometer in combination with the pre-filter}

\begin{figure}
\begin{center}
\centerline{\hspace*{-0.05\textwidth}
\includegraphics[width=0.6\textwidth,clip=]{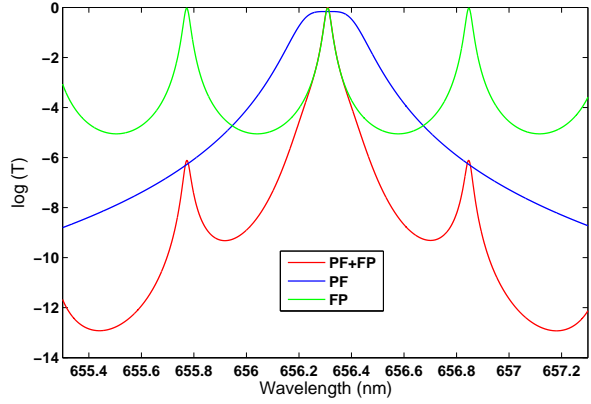}
           }
  \vspace{-0.05\textwidth}    
               
\end{center}
\caption{The red curve shows the transmission profile of the FP in combination with a pre-filter.
The green and blue curves show the transmission profiles of the FP and pre-filter, respectively.}
\label{fig:nbi_trans}
\end{figure}

In the NBI setup, a single FP is positioned in tandem with the order sorting interference filter 
(see Fig.~\ref{fig:1}). If two such optical filters are placed in series
then the resulting instrumental profile $I(\lambda)$ is obtained by \citep{Cavallini06,Ravindra10},

\begin{equation} \label{equ:6}
I(\lambda) = I_{PF}(\lambda) \cdotp I_{FP}(\lambda)
\end{equation}
where $I_{PF}(\lambda)$ and $I_{FP}(\lambda)$ are the transmission of the order sorting filter and FP, respectively.
In estimating the transmission profile of the NBI we have used the spacing
of the FP as 0.4~mm with a plate reflectivity of 85\% and the FWHM of the interference filter is 2.3~\AA.
The FP and pre-filter are placed in tandem in a collimated beam.
The red curve in Fig.~\ref{fig:nbi_trans} shows the theoretically estimated transmission profile of 
the FP in combination with a pre-filter. As the pre-filter is an interference filter,  we adopted the profile of 
the pre-filter to be Lorentzian \citep{Cavallini06}.
The green and blue curves show the transmission profiles of the FP and pre-filter, respectively. From the 
combined transmission profile it is clear that the out of band rejection is large outside the pre-filter profile.

\begin{figure}
\begin{center}
\centerline{
\includegraphics[width=0.8\textwidth,clip=]{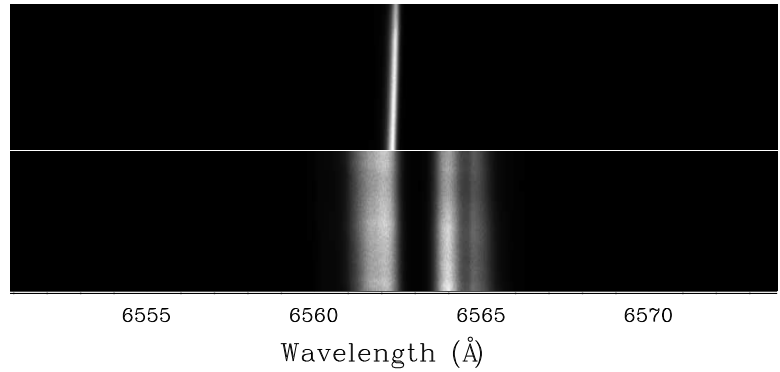}
           }
  \vspace{-0.05\textwidth}    
               
\end{center}
\caption{Top panel shows the channel spectrum of the FP in combination with the pre-filter.
Bottom panel shows the observed solar spectrum using only the pre-filter.} 
\label{fig:nbi_trans_exp}
\end{figure}

We also conducted an experiment to see the out of band rejection. 
This was done by recording the FP channel spectra by keeping the pre-filter and FP in a collimated beam configuration 
in front of the spectrograph.
The channel spectrum is shown in the top panel of Fig.~\ref{fig:nbi_trans_exp}. The bottom panel shows 
the observed solar spectrum using the pre-filter when the FP 
was removed from the optical path. The position of the channel spectrum can be shifted
by changing the plate separation of the FP and obtaining 
images at different wavelength positions.

\section{Preliminary Observations}
\label{s:observation}
\begin{figure}    
 \centerline{\hspace*{-0.015\textwidth}
               \includegraphics[width=0.8\textwidth,clip=]{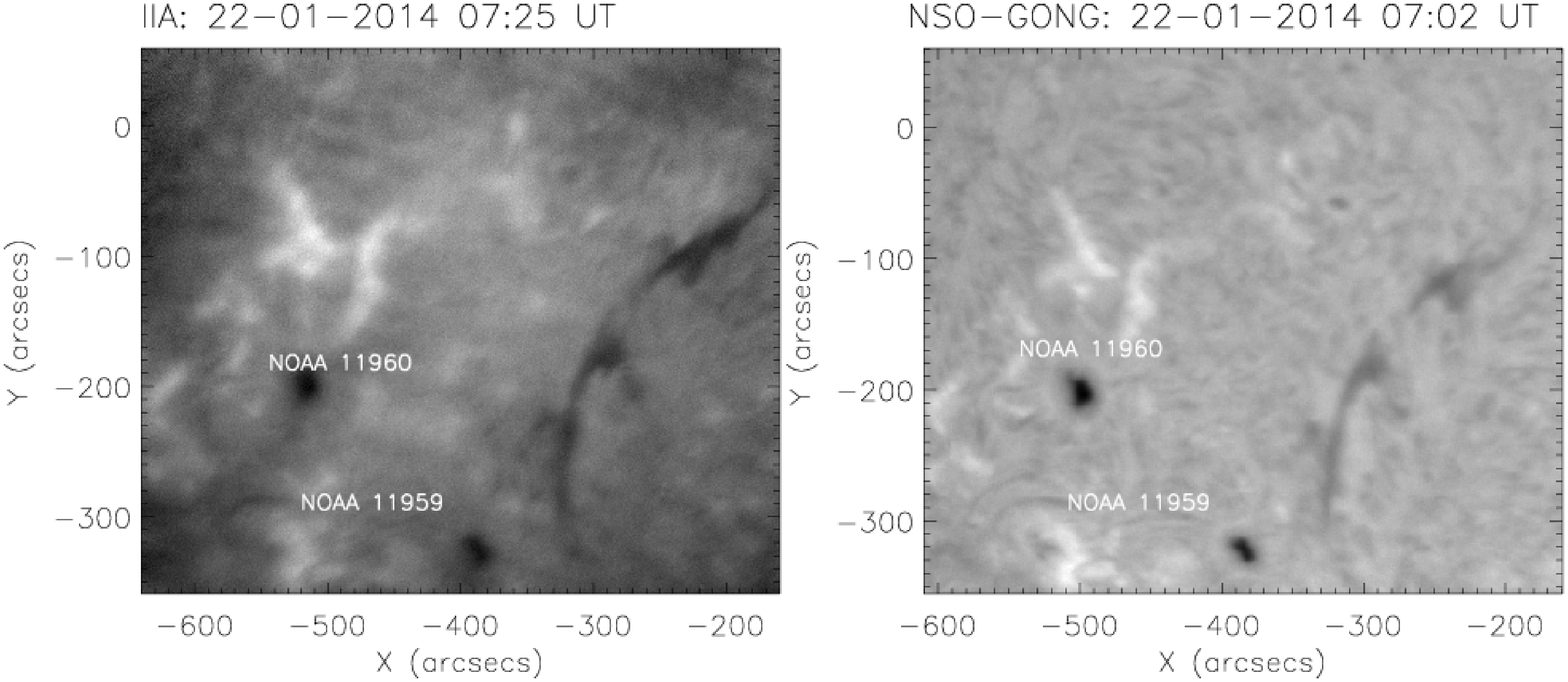}               
              }
    
      \vspace{0.01\textwidth}    
      
        \centerline{\hspace*{-0.015\textwidth}
               \includegraphics[width=0.8\textwidth,clip=]{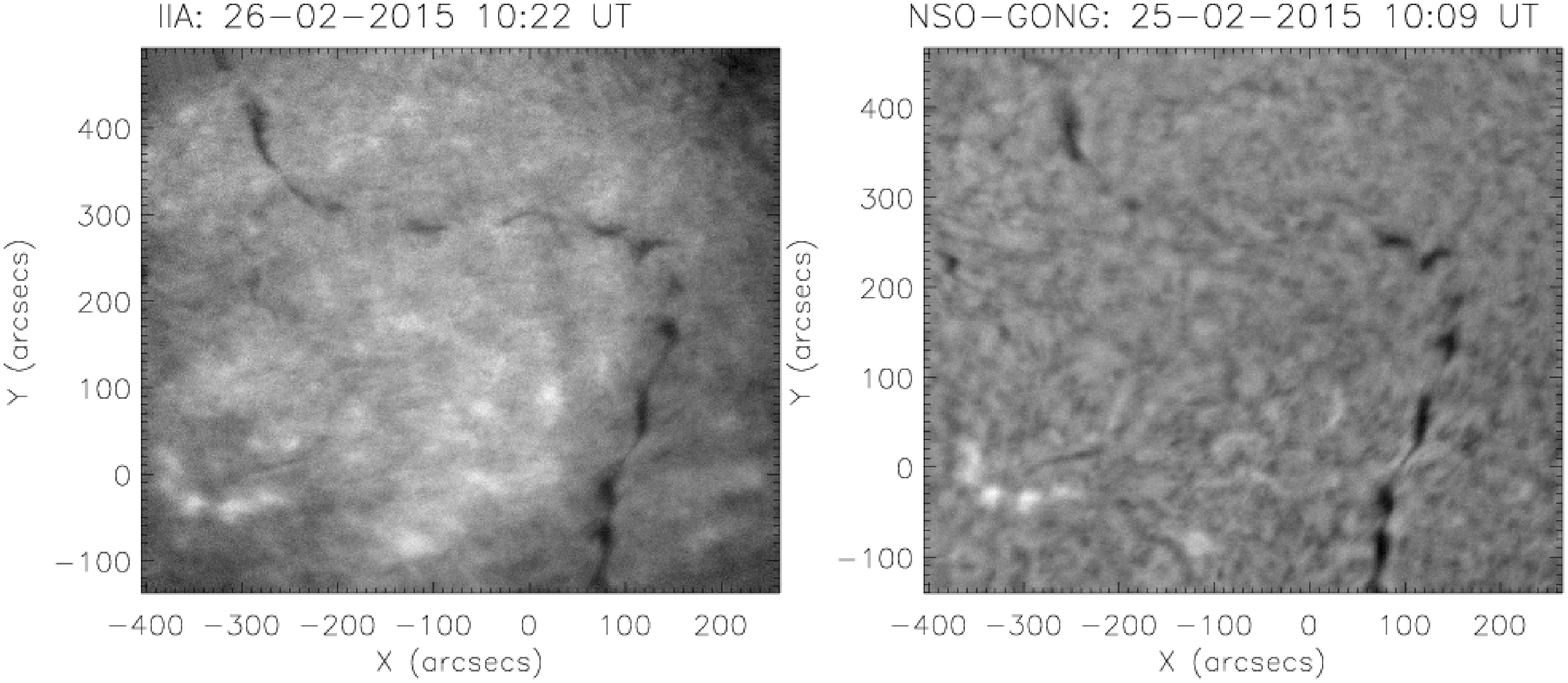}     
              }
              \vspace{0.01\textwidth}    
      
        \centerline{\hspace*{-0.015\textwidth}
               \includegraphics[width=0.8\textwidth,clip=]{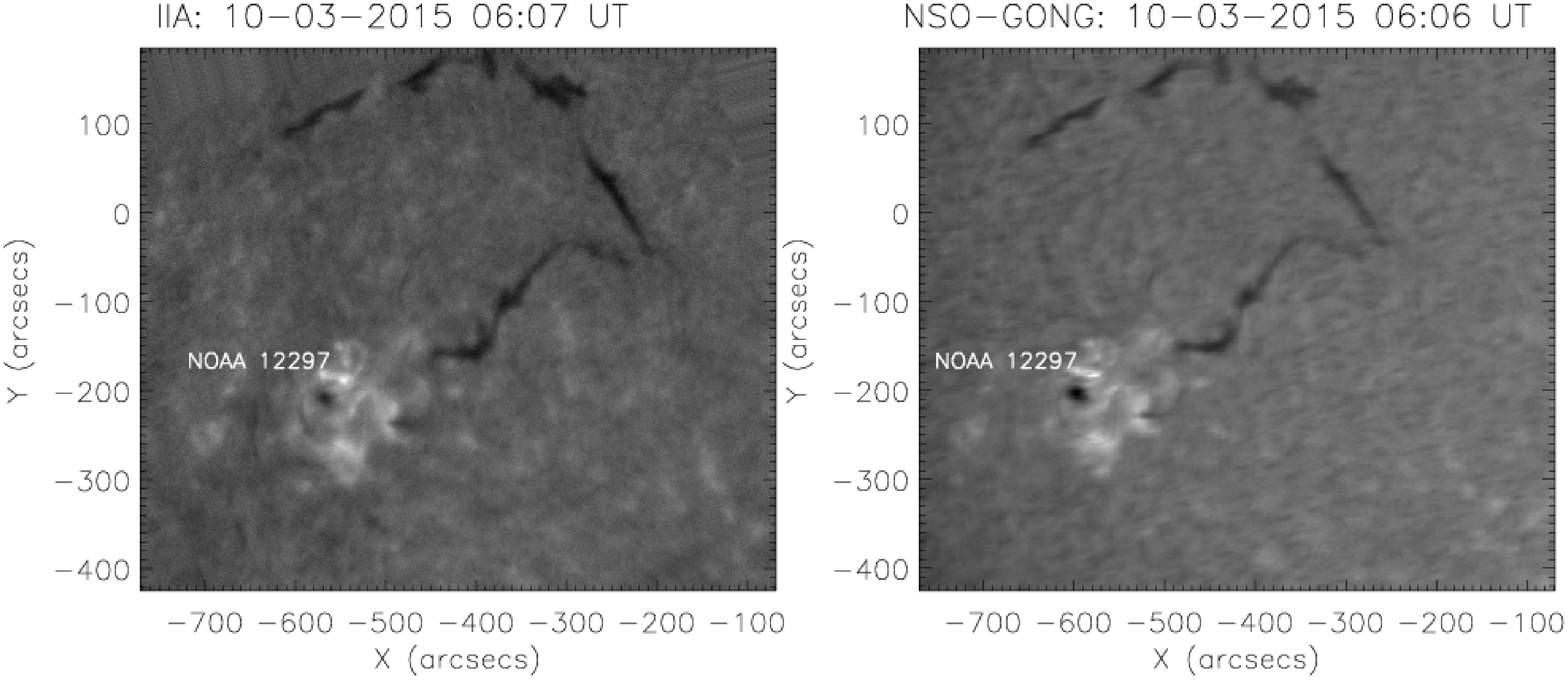}     
              }        
       \vspace{0.05\textwidth}    
     
\caption{Left: top, middle and bottom panels show the observed filaments at different times using NBI at the H$\alpha$
line core positions.
Right: top, middle and bottom panels show the observed filaments at the corresponding dates from NSO-GONG.}

   \label{fig:obs1}
   \end{figure}

After testing the FP, we used it as a main component for developing the NBI.   
The schematic layout of the NBI is shown in Fig.~\ref{fig:1}. The NBI system 
is setup in the laboratory of IIA, Bangalore campus.
The pre-filter, centered on 6562.8~\AA,  was employed
ahead of the FP to select the desired wavelength channel for observations.
A series of images was obtained by tuning the plate separation of FP
across different wavelength positions on the H$\alpha$ spectral line. 
Fig.~\ref{fig:obs1} shows the images taken on different days in the line center at 6563 ~\AA. 
Top-left panel shows the image of active regions NOAA 11960 and NOAA 11959 and a quiet filament taken using the NBI.
These two sunspots (NOAA 11960 and NOAA 11959) were visible on the southern hemisphere of the 
Sun from 2014 January 19 to January 29 and they were associated with a series of C-class flares.
We have observed these sunspot regions on 2014 January  22 and they are shown 
in Fig.~\ref{fig:obs1} (top-left). This image is compared with the image taken from the
\textit{Global Oscillation Network Group} (GONG) observed in the $H\alpha$ 
wavelength taken close to the time of our observation (top-right). Clearly, both 
images show the sunspots and filament structure. 
The contrast in the filament is better in the FP based NBI image compared to that from GONG. This
is because of the very narrow band of the NBI filter compared to  GONG, which is about 0.4~\AA~\citep{Harvey11}.
Middle-left and right images in Fig.~\ref{fig:obs1} show the image taken using the NBI and GONG respectively
for a quiet region filament on 2015 February 26. It was a very long filament and the size was about 750 Mm,
which covered almost half the diameter of the Sun. This filament was visible on the solar disk 
from 2015 February 22  to March 2.
We observed part of the filament on 2015 February, 26 (shown in  Fig.~\ref{fig:obs1} (middle-left)).
Bottom-left and right panel in Fig.~\ref{fig:obs1} show the image of active region NOAA 12297 
on 2015 March 10 using the NBI and GONG, respectively. In both images, one can a see
dark sunspot surrounded by bright patches
and a large quiet filament structure. This sunspot appeared on the solar disk 
at S17~E53 on 2015 March 8 and it was visible
on the solar disk for another 10 days. This sunspot region was associated with a
series of C and M-class flares. We observed this
region for a few days. In the NBI images we see more structures compared to the GONG images.

\begin{figure}    
 
   \centerline{\hspace*{0.015\textwidth}
               \includegraphics[width=0.85\textwidth,clip=]{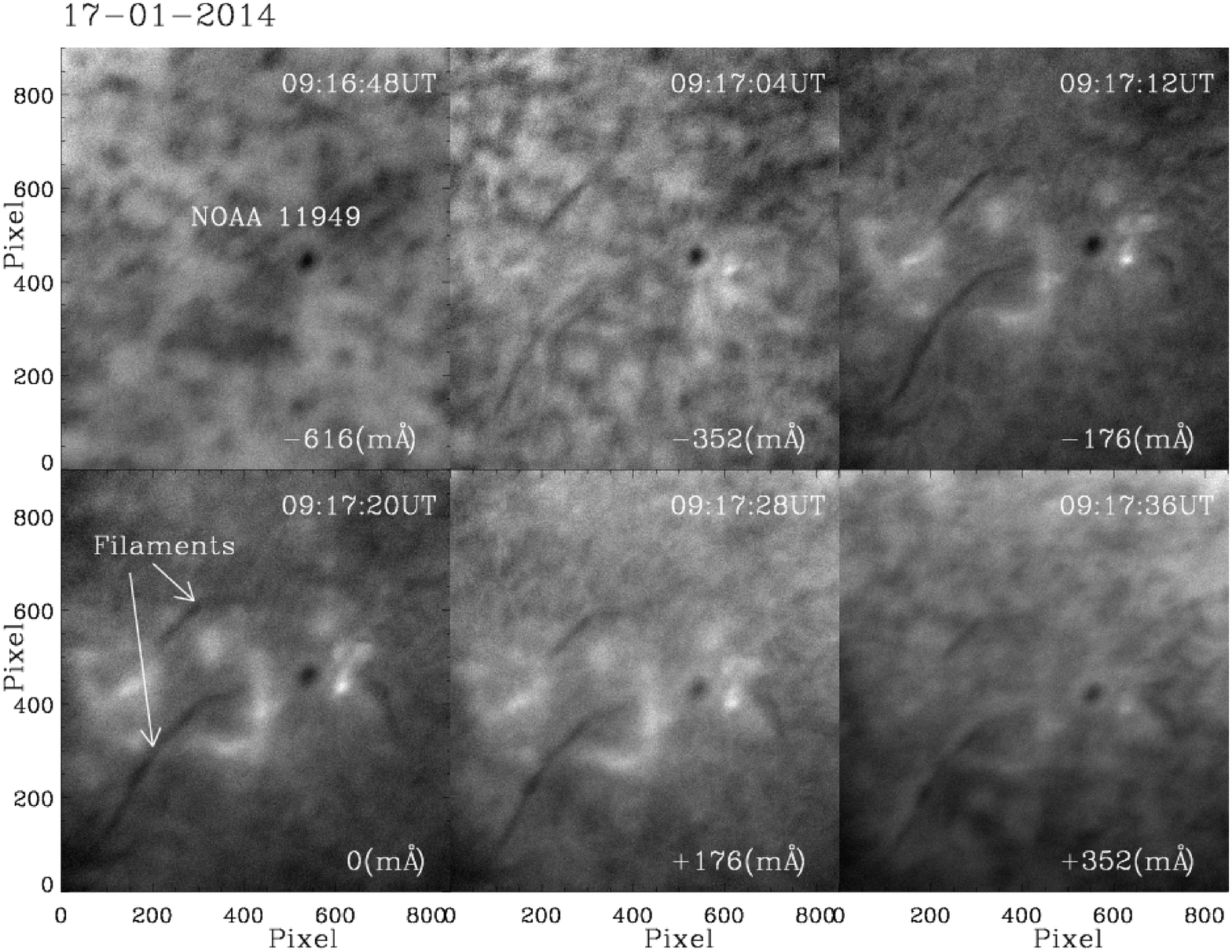}               
              }
    \vspace{-0.07\textwidth}   
   \centerline{\hspace*{0.015\textwidth}
               \includegraphics[width=0.85\textwidth,clip=]{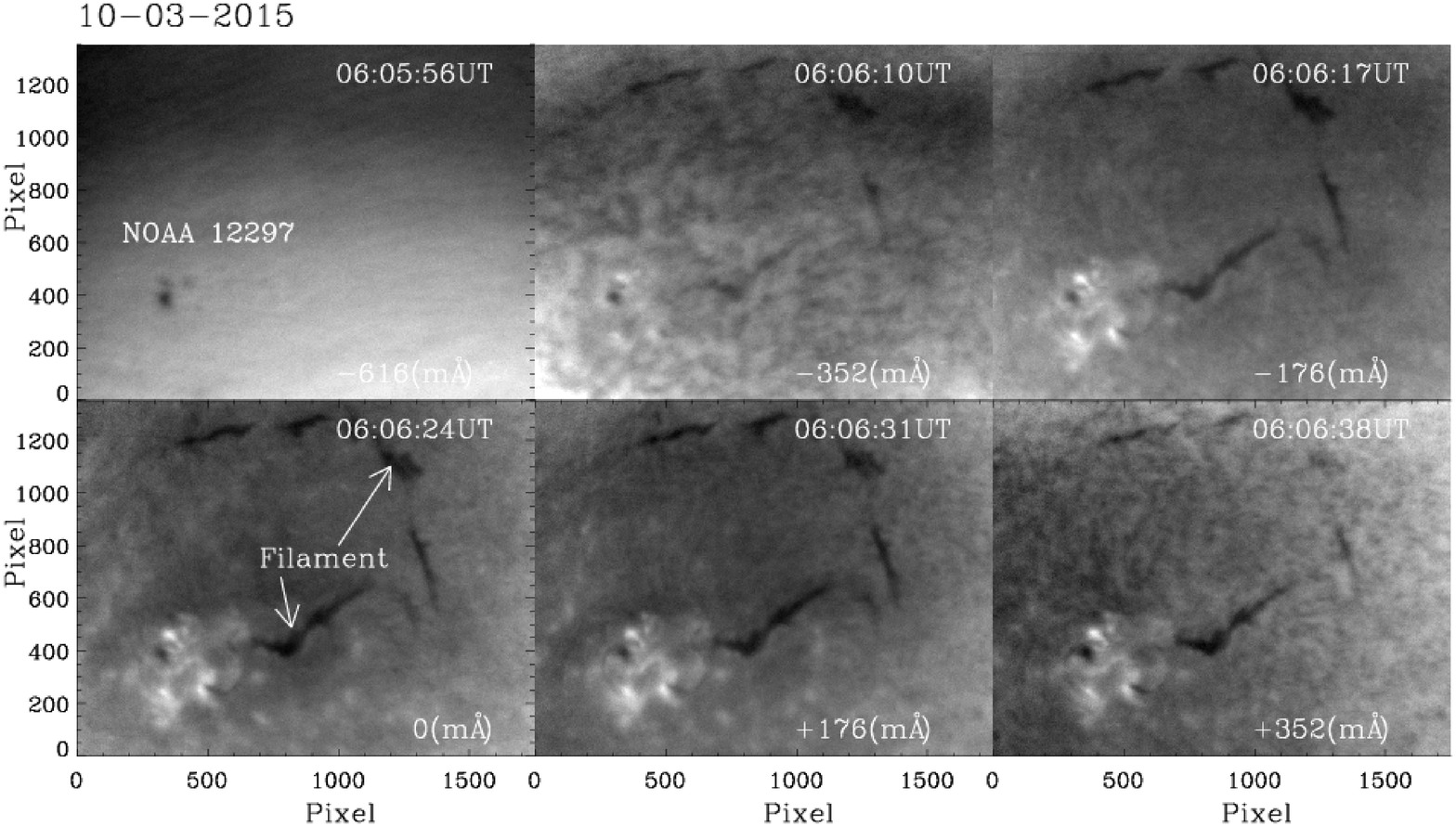}               
              }    
      \vspace{0.05\textwidth}    
                 
\caption{Images obtained at the different wavelength positions of the H$\alpha$ 6563~\AA~ line profile.
The images have spectral resolution 0.37``/pixel. Top panel images were taken on  2014 January 17 for 
the active region NOAA 11949 and bottom panel images were taken on 2015 March 10 of the active region NOAA 12297.
}

   \label{fig:obs2}
   \end{figure}

Images acquired at different wavelength positions on the line profile are shown in Fig.~\ref{fig:obs2}. 
These images were obtained by scanning the H$\alpha$ spectral line starting 
from 792 m\AA~on the blue wing of the spectrum to 616 m\AA~on the red wing away from the H$\alpha$
line center, with a step of 88~m\AA. 
After the dark current correction and flat fielding ~\citep{Chae04}, the images are shown here. 

\begin{figure}
\begin{center}
\centerline{\hspace*{-0.05\textwidth}
\includegraphics[width=0.6\textwidth,clip=]{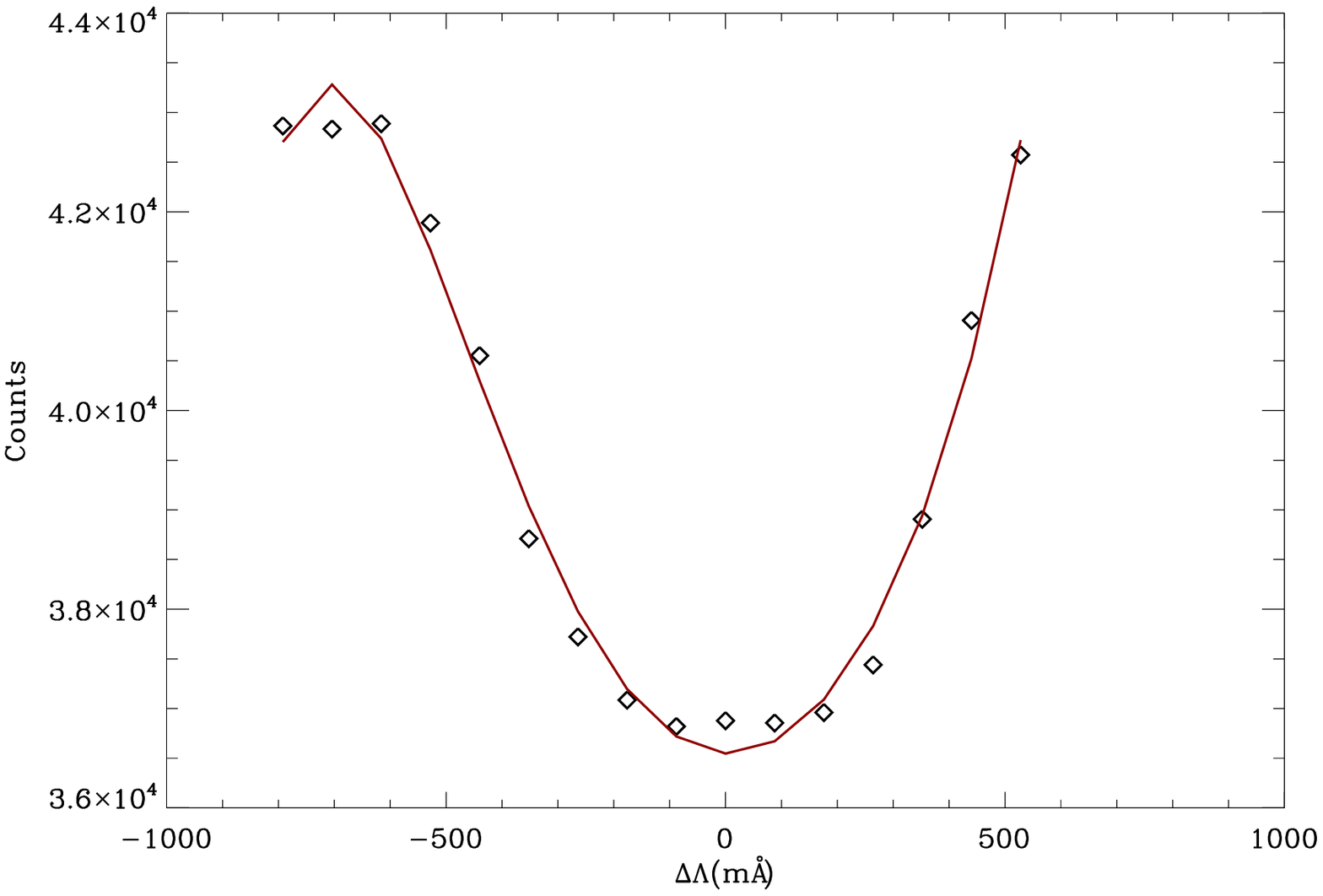}
           }
  \vspace{-0.05\textwidth}    
               
\end{center}
\caption{Line profiles  at  6563~\AA~ constructed from observations. 
The line profile is scanned at 18 wavelength positions from
-792~m\AA~to +616~ m\AA~ with a step size of 88~m\AA. }

\label{fig:scan}
\end{figure}

In Fig.~\ref{fig:obs2}, top panel, we show the images obtained at different 
wavelength positions for the active region NOAA 11949. 
This sunspot was located at S17~W46 on 2014 January 17. Two filament structures are visible 
in the line center and close to line center images, as shown by the two arrows. 
It can be noted that the quality of these images is
affected by the poor seeing conditions at Bangalore. The image is acquired with an FOV 12.6$^{\prime}$ and has
pixel resolution of 0.37$^{\prime\prime}$. 
We also performed a spectral line scan for another active region NOAA 12297 on 2015 March 10. 
This active region was located
at S16~E26. The observed images at different wavelength
positions are shown in Fig.~\ref{fig:obs2} (bottom panel).
We have constructed the H$\alpha$ spectral line profile from the scanned images after dark current correction and flat fielding. 
We calculated the mean intensity of 100$\times$100 pixel$^2$ from the quiet part of the Sun to retrieve the line profile.
Fig.~\ref{fig:scan} shows the obtained mean intensity at different wavelength positions. In addition, we fitted a
higher order polynomial to construct the H$\alpha$ absorption line profile. The red curve shows the polynomial 
fitting to the mean intensity.
One can see that the line profile could not be retrieved completely. This could be due to the 
small passband of the pre-filter.

\subsection{Dopplergram}
To study the line of sight (LOS) velocity of solar features, Doppler shifts of the spectral lines are regularly used. 
The Doppler shift of the spectral line $\lambda$ for a defined velocity v is calculated by 
using the relation 

\begin{equation} \label{equ:7}
\frac{\Delta \lambda}{\lambda} = \frac{v}{c}
\end{equation}
where c is the speed of light.  
In order to compute the Doppler velocity map, the line center
and computed shift of wavelength from the line center are required.
This can only be done if the simultaneous 
spectral line data for the same region are available (one from the spectrograph and other from NBI).

Alternatively, using an FP based tunable NBI one can make the Doppler velocity maps by using the red and blue wing
intensity images of a chosen spectral line. Intensity at a fixed wavelength point in the
blue and red wing of any spectral line varies depending on the Doppler shift. 
Hence, from the intensity difference between the red- and blue-wing images, one can estimate the Doppler shift 
and hence the Doppler velocity at a fixed wavelength position. For a small line profile shift, 
one may assume the solar spectral line is linear as well as symmetric, so
Doppler velocities can be estimated by using the following relation:

\begin{equation} \label{equ:8}
v = K \frac{I_{r}-I_{b}}{I_{r}+I_{b}}
\end{equation}

where, $I_{r}$ and $I_{b}$ are the intensity images at red wing and blue wing of a spectral line. 
K is a calibration constant which
depends on the chosen spectral line and the spectral resolution \citep{Stix04,Rimmele04,Padinhatteeri10}. 
The denominator in eq.~\ref{equ:8} corrects the intensity variations. $I_{r}$ and $I_{b}$
can be expressed as 

\begin{eqnarray} \label{equ:10}
 I_{r}= I + \frac{dI}{d\lambda}\Delta\lambda \nonumber\\
 I_{b}= I - \frac{dI}{d\lambda}\Delta\lambda
\end{eqnarray}

Where \textit{I} is the photometric signal at the spectral line core position and \textit{dI} represents the variations 
of signal with wavelength variation \textit{d$\lambda$} from the central wavelength $\lambda$. 
By combining eq.~\ref{equ:7} and eq.~\ref{equ:10}, we can get

\begin{eqnarray} \label{equ:11}
\frac{I_{r}-I_{b}}{I_{r}+I_{b}} = \frac{\frac{dI}{d\lambda}}{I}\frac{\lambda v}{c}\nonumber\\
v=\frac{Ic}{\lambda \frac{dI}{d\lambda}}\frac{I_{r}-I_{b}}{I_{r}+I_{b}}
\end{eqnarray}
By comparing eq.~\ref{equ:8} and eq.~\ref{equ:11}, one can obtain the calibration constant \textit{K} as

\begin{equation} \label{equ:12}
K = \frac{c}{\lambda} I \frac{d\lambda}{dI}
\end{equation}
If we plot the variations of $\frac{dI}{I}$ against $\frac{d\lambda}{\lambda}$, the slope of the graph will equal 
$\frac{c}{K}$. To estimate the calibration constant \textit{K}, we have taken an H$\alpha$ spectral profile 
from the solar atlas \textit{BASS 2000} (\textit{http://bass2000.obspm.fr/}). We used the
238~m\AA~FWHM of our NBI. The spectral line is convolved with the instrumental profile.
The convolved spectral line is then shifted by an
amount  $d\lambda$ and the normalized intensity difference ($\frac{dI}{I}$) is calculated. 
We repeated this procedure for a range of velocities. 

\begin{figure}
\begin{center}
\centerline{\hspace*{-0.05\textwidth}
\includegraphics[width=0.6\textwidth,clip=]{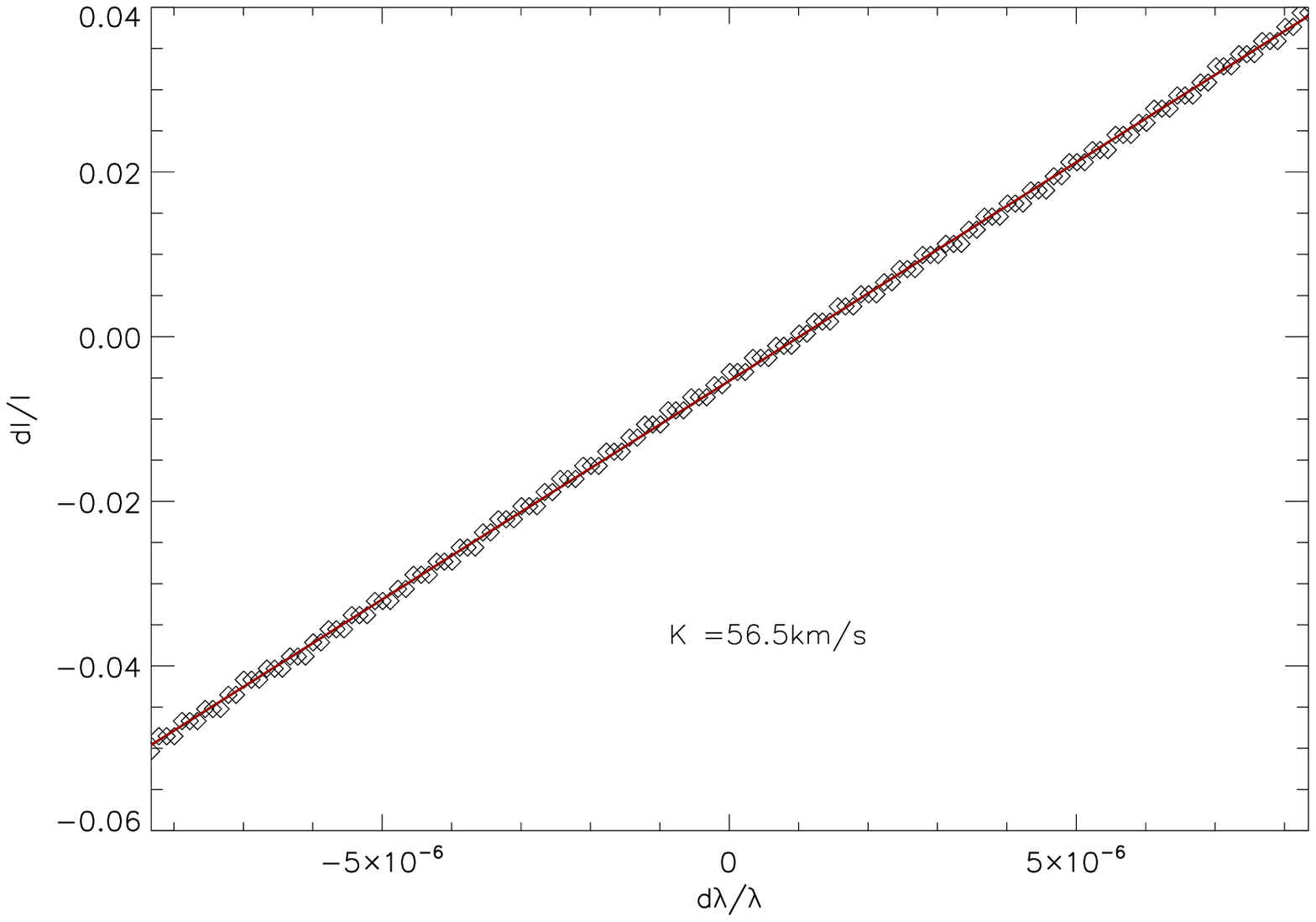}
           }
  \vspace{-0.05\textwidth}    
               
\end{center}
\caption{The normalized intensity difference versus
$\frac{d\lambda}{\lambda}$ curve estimated for the spectral line H$\alpha$ 6562.8~\AA. The red line shows the straight line
fit to the estimated values.}

\label{fig:calibration_constant}
\end{figure}

Fig.~\ref{fig:calibration_constant} 
shows the variations of $\frac{dI}{I}$
as a function of $\frac{d\lambda}{\lambda}$. The linear part of the curve is fitted with a straight line (red line)
and \textit{c/slope} provides the values of calibration constant \textit{K}. The calculated 
value of \textit{K} is 56.5$\pm$0.1km/s.


 \begin{figure}    
 
   \centerline{\hspace*{0.015\textwidth}
               \includegraphics[width=1.0\textwidth,clip=]{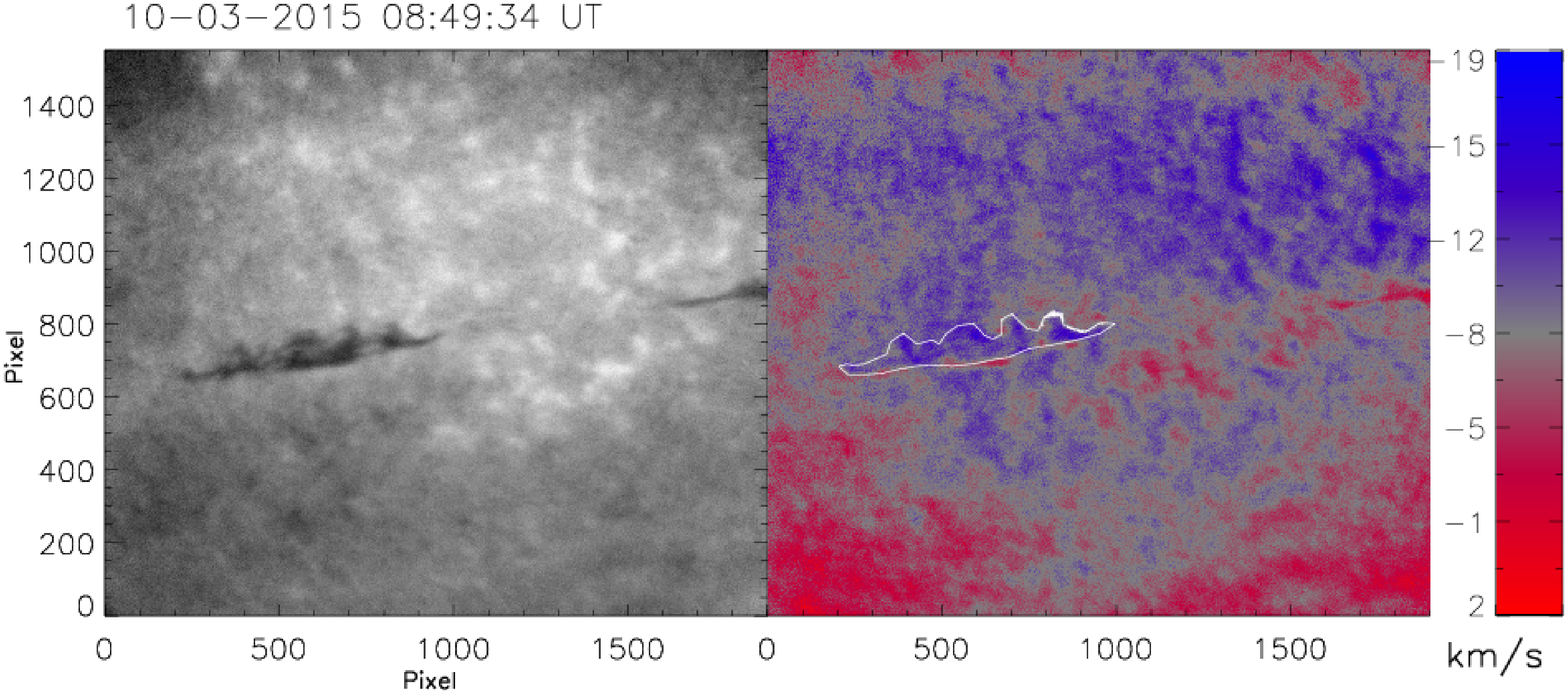}               
              }
    \vspace{0.01\textwidth}   
   \centerline{\hspace*{0.015\textwidth}
               \includegraphics[width=1\textwidth,clip=]{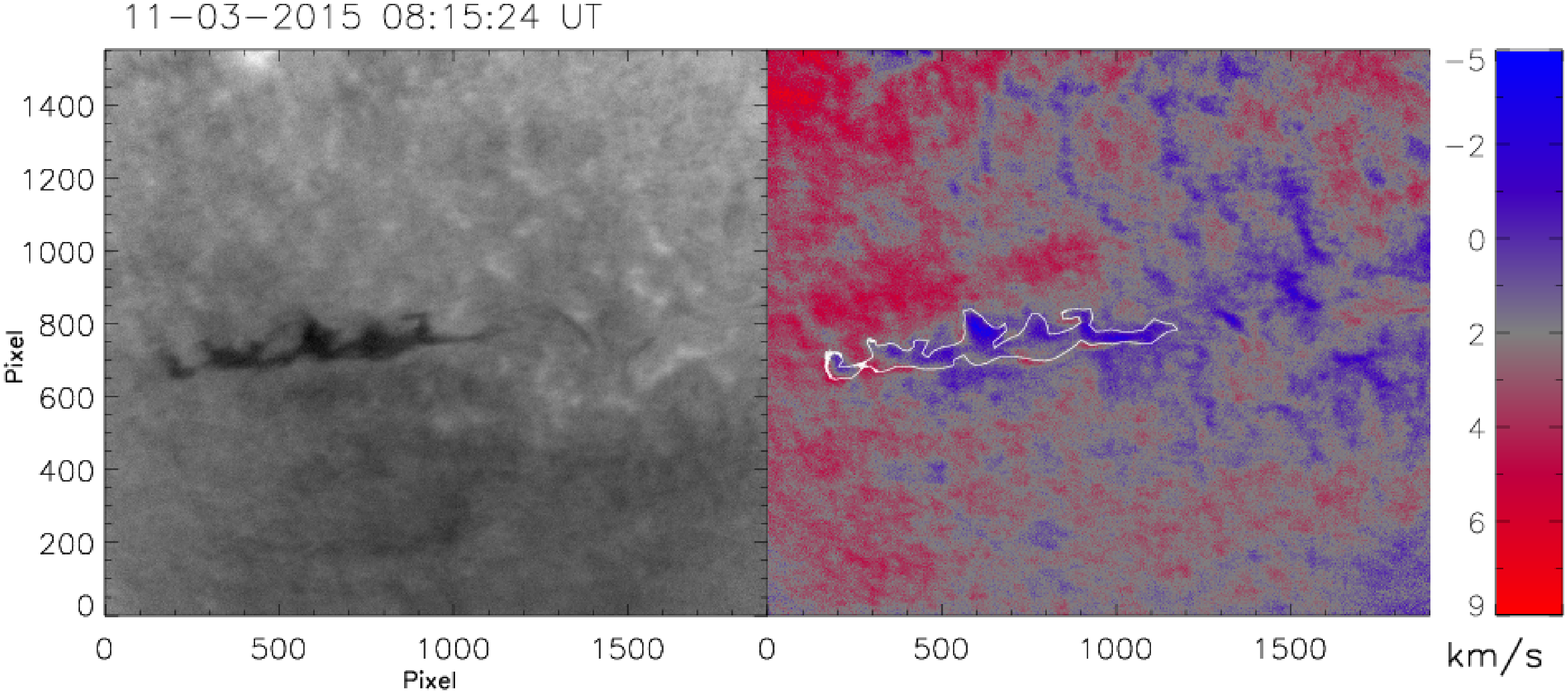}               
              }       
      \vspace{0.05\textwidth}    
                 
\caption{Left: top and bottom panels show the images taken at the H$\alpha$ line core position (6562.8~\AA) for
the quiet Sun region on 2015 March 10 and 11, respectively. Right: Images show the corresponding generated dopplergrams 
at the chromospheric height. The filament contour is overlaid on the dopplergram.
}

   \label{fig:dopplergram1}
   \end{figure}

 \begin{figure}    
 
   \centerline{\hspace*{0.015\textwidth}
               \includegraphics[width=1.0\textwidth,clip=]{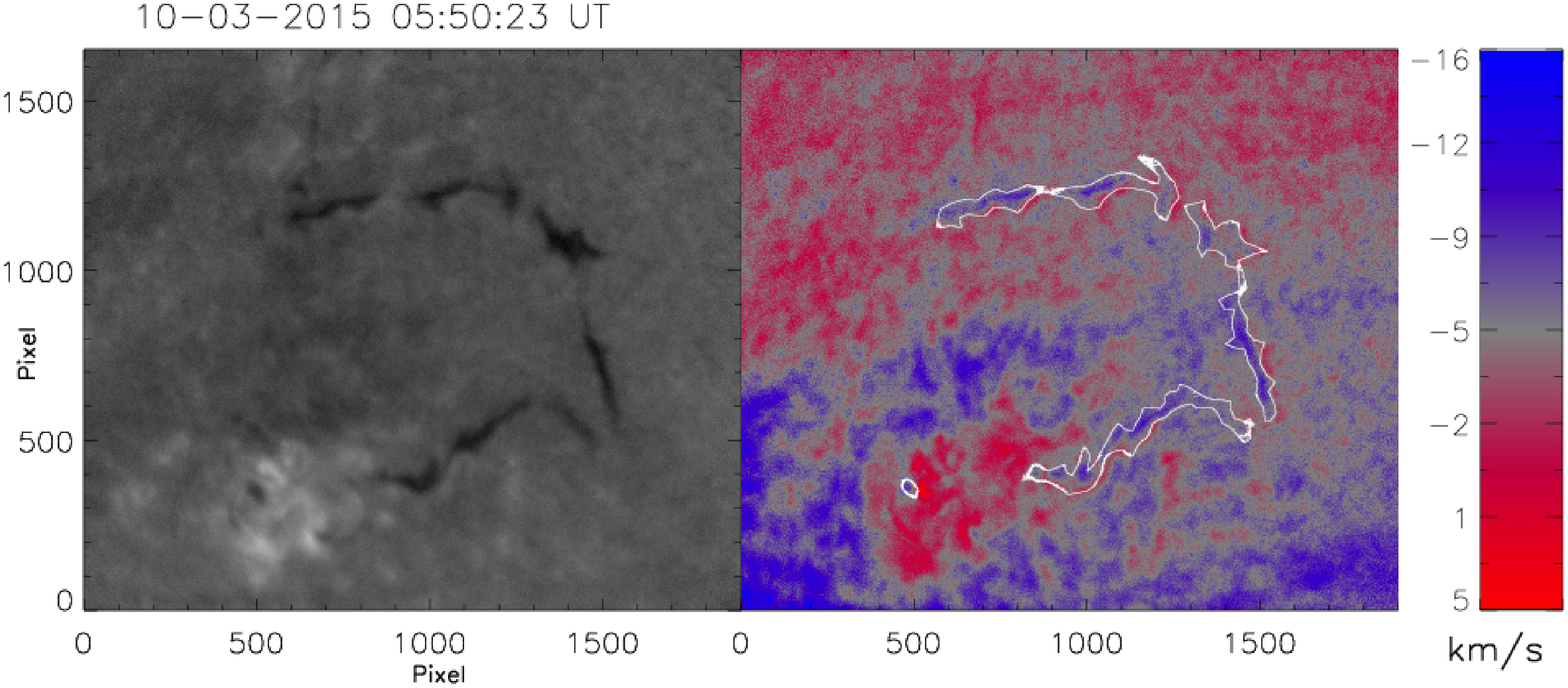}               
              }
    \vspace{0.01\textwidth}   
   \centerline{\hspace*{0.015\textwidth}
               \includegraphics[width=1\textwidth,clip=]{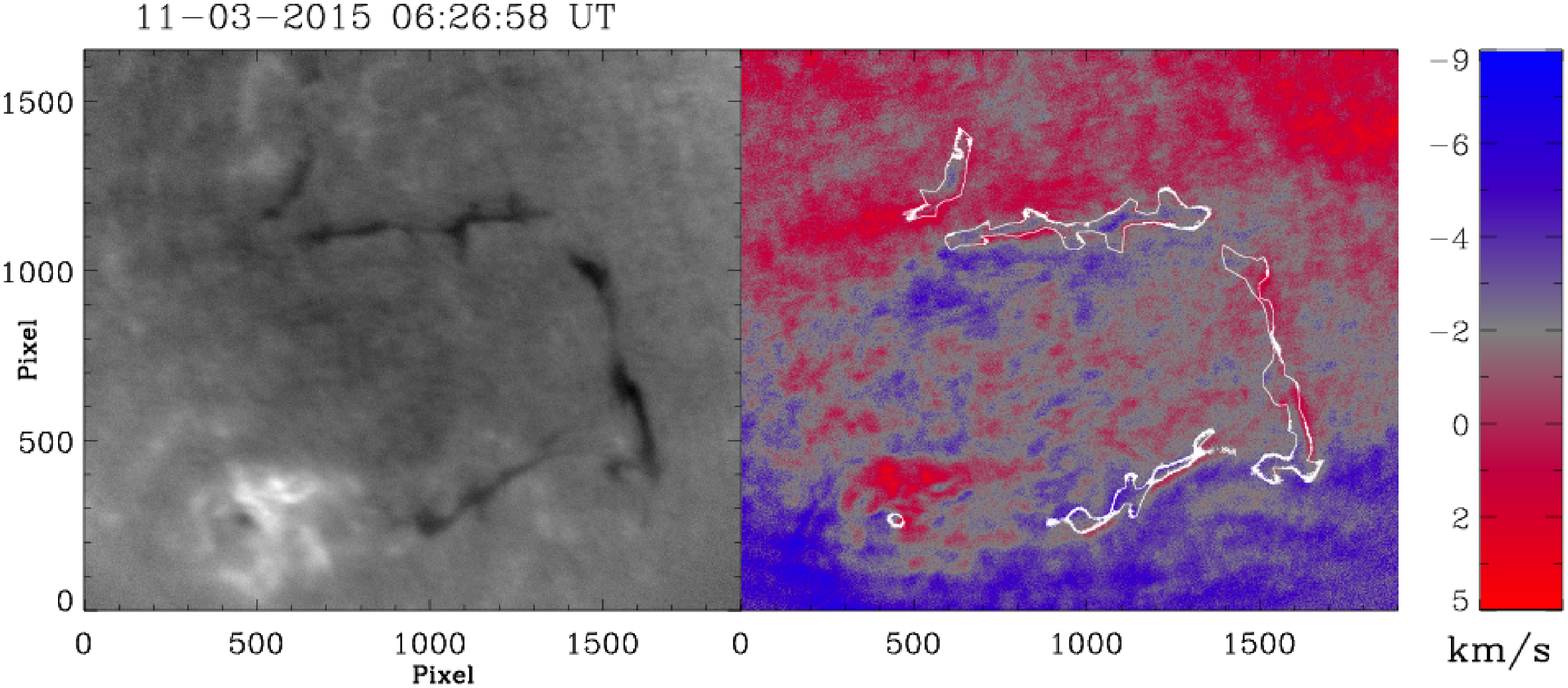}               
              }       
      \vspace{0.05\textwidth}    
                 
\caption{Left: top and bottom panels show the images taken at the H$\alpha$ line core position 
(6562.8~\AA) for the active region
NOAA 12297 on 2015 March 10 and 11, respectively. Right: Images show the corresponding dopplergrams 
at the chromospheric height. The filaments and sunspot contours are overlaid on the dopplergram.
}

   \label{fig:dopplergram2}
   \end{figure}

To generate the dopplergrams, we carried out observations using our NBI. We obtained the 
blue and red wing images at $\pm$176 m\AA~ wavelength positions separately away from the line center (6562.8~\AA) of the
H$\alpha$ spectral line. The dopplergrams have been generated using eq.~\ref{equ:8}.
Fig.~\ref{fig:dopplergram1}, right side, shows the obtained dopplergram for a small FOV at chromospheric height for a quiet-Sun region.
The corresponding images on the left side were taken at the H$\alpha$ line core 
wavelength position (6562.8~\AA) for the same region.
One can see a quiet filament structure is visible in the image. The white contour oulining the 
filament is overlaid on the dopplergram to
mark its position. This filament was visible on the southern hemisphere of the Sun from 2015 March 9 to 16. 
On March 10, the location of the filament was at (-400$^{\prime\prime}$, -500$^{\prime\prime}$) and it was 180 Mm long.
Along the spine of the filament we observed a blue shift. The maximum and minimum LOS velocities are shown by the color 
bar in the images. By analyzing the doppler velocity maps, we found that redshifts
appeared near the east end of the filament.
We also generated the dopplergram for the active region NOAA 12297 and its surrounding region
on 2015 March 10 and 11. On the March 10, the active region was located at S19~E26. The angle
subtended by the solar disk center and the active region was 32$^{\circ}$. The maximum and minimum velocities along the 
LOS in the FOV of our observation are $+$5 km/s and $-16$ km/s, 
respectively, as shown in Fig.~\ref{fig:dopplergram2}. In the periphery of the filament of 
the same dopplergram, one can see
blue and red-shift of LOS velocity, as a similar type of velocity map was reported by \cite{Joshi15}. 
In that case, the true maximum and minimum velocities along the LOS in the spine of the filament were 
about $+$5.8 km/s ($=$ 5/cos 32$^{0}$) and $-$18.8 km/s ($= -$ 16/cos 32$^{0}$), respectively. 
We have observed the same filament for two more days and no significant 
change in the filament structure was observed. Hence from our observation, we can infer that 
the observed blueshift and redshift at the periphery of the filament were the downflows and 
upflows, respectively across the filament spine.

\section{Discussions}\label{s:discussion}

We have developed a single FP based NBI at the Indian IIA, Bangalore. We tested the parameter value for each 
of the optical components in a laboratory and compared them with the specified parameters provided by the vendor. 
Most of the observed parameters compare well with the vendor's specifications with a few deviations in some cases.
This is mostly because of the reduced reflectivity of our FP over time. We tested the NBI 
system in a laboratory and finally we made several days of solar observations. We also scanned the H$\alpha$ 
line profile and obtained images at each of the wavelength positions. We compared our images with the GONG H$\alpha$ images
and found the features match well in both the images.

As a step further we also constructed  dopplergrams by making the observations in the blue and red wing of the H$\alpha$ spectral line.
The H$\alpha$ line center forms in the upper chromosphere and the obtained dopplergrams represents the velocity at the
chromospheric level.  At present there are no chromospheric 
dopplergrams available on regular basis. Hence we cannot make a detailed comparison of the 
developed dopplergrams with other
data. However, we find some similarity in the filament region with dopplergrams produced by \cite{Joshi15}. 
In thae future, we would like to make a comparison with the 
dopplergrams in the H$\alpha$ from the Kodaikanal Solar Observatory using the Lyot filters. 

The spectral resolution of the single FP based NBI system is small. Using two or more FPs in tandem one can increase 
the spectral resolution of the imaging system as has been done in IBIS \citep{Cavallini06,Reardon08} and TESOS \citep{Tritschler04}. 
With this we can also increase the pass band 
of the prefilter and hence we will be able to scan the full spectral line profile. At the IIA
we are designing one such instrument for the National Large Solar Telescope (NLST: \cite{Hasan10}) which will be installed
at one of the high altitude Himalayan sites. Using this we will be able to study the dynamics of the filaments both in the 
photosphere and chromosphere nearly simultaneously at very high spatial resolution (0.05$^{\prime\prime}$).

\normalem
\begin{acknowledgements}

We would like to thank Mr. Periyanayagam and Mr. Thimmaih N. for their help in making the mechanical items for the instruments.
We also acknowledge Dr. Shibu K. Mathew  for helping us in the design and 
fabrication of the temperature-controller for the pre-filter oven. 
We thank the referee for insightful comments which helped us to improve the content in the
manuscript.

\end{acknowledgements}
  

\end{document}